\documentclass[aps,epsf,preprint,nofootinbib]{revtex4}
\usepackage{pifont}
\usepackage{amssymb}
\usepackage{amsmath}
\usepackage{delarray}
\usepackage{rotating}
\usepackage[usenames]{color}

\newcommand{\nslash}{\kern 0.2 em n\kern -0.50em /}

\newcommand{\beq}{\begin{eqnarray}}
\newcommand{\eeq}{\end{eqnarray}}

\def\bq{\begin{eqnarray}}
\def\eq{\end{eqnarray}}
\def\bq{\begin{equation}}

\def\roughly#1{\mathrel{\raise.3ex\hbox{$#1$\kern-.75em
\lower1ex\hbox{$\sim$}}}}

\begin{document}

\preprint{\hfill\parbox[b]{0.3\hsize}
{ }}

\def\bra{\langle }
\def\ket{\rangle }

\title{Analyzing the Boer-Mulders function within different quark models}

\author{A. Courtoy$^1$\footnote{E-mail address: aurore.courtoy@uv.es},
S. Scopetta$^{2}$\footnote{
E-mail address: sergio.scopetta@pg.infn.it},
V. Vento$^{1}$\footnote{E-mail
address: vicente.vento@uv.es}
}
\affiliation
{\it
(1)
Departament de Fisica Te\`orica, Universitat de Val\`encia
\\
and Institut de Fisica Corpuscular, Consejo Superior de Investigaciones
Cient\'{\i}ficas
\\
46100 Burjassot
(Val\`encia), Spain
\\
(2)
Dipartimento di Fisica, Universit\`a degli Studi di Perugia, and
\\
INFN, sezione di Perugia, via A. Pascoli
06100 Perugia, Italy
}

\begin{abstract}

A general formalism for the evaluation of time reversal
odd parton distributions
is applied here to calculate the Boer-Mulders function.
The same formalism when applied to evaluate
the Sivers function  led to results which  fulfill the Burkardt
sum rule quite well.
The calculation here has been performed for two different models of proton structure:
a constituent quark model
and the MIT bag model.
In the latter case, important differences are found
with respect to
a previous evaluation in the same framework,
a feature already encountered in the calculation of the Sivers function.
The results obtained are consistent with the present wisdom, i.e.,
the contributions for the $u$ and $d$ flavors turn
out to have the same sign, following the pattern
suggested analyzing the model independent features of the
impact parameter dependent
generalized parton distributions.
It is therefore confirmed that
the present approach is
suitable for the analysis
of time reversal odd distribution functions.
A critical comparison between the outcomes of the
two models, as well as between the results of the calculations
for the Sivers and Boer-Mulders functions, is also carried out.

\end{abstract}

\maketitle

\section{Introduction}

The study of the transverse polarization of quarks in the nucleon,
one of its less known features
(for a review, see, e.g.,
Ref. \cite{bdr}), is progressing fast, since the expected
new data  from ongoing
experiments 
are motivating a strong theoretical
activity (for recent developments, see Ref. \cite{ferrara}).
The present work aims to contribute
to the understanding of the transverse polarization of quarks
by evaluating, using different models
of the proton structure,
the Boer-Mulders function \cite{Boer:1997nt}. We will use
 a theoretical scenario \cite{Courtoy:2008vi,Courtoy:2008dn}
recently developed for the calculation of the
Sivers function \cite{sivers}, which has reproduced
its main features, such as the sign, the Burkardt sum rule
\cite{Burkardt:2004ur},
and the general trend of the parameterizations
extracted from the available data \cite{ans}.

The Boer--Mulders function $h_1^\perp$
describes the
number density of transversely polarized quarks
in an unpolarized target.
As
the Sivers function, $f_{1T}^\perp$,
describing the number density of unpolarized quarks in
a transversely polarized target,
the Boer--Mulders function is
a Transverse Momentum Dependent (TMD) PD and
it is a time-reversal odd object (T-odd) \cite{bdr}. For this reason,
for several years, it was believed to vanish due to time reversal invariance.
However, this argument was invalidated, initially
in the case of the Sivers function, by a calculation
in a spectator model \cite{brohs}, following the observation
of the existence of leading-twist
Final State Interactions (FSI) \cite{brodhoy}.
The current wisdom is that non-vanishing T-odd functions
are generated by the gauge link in the definition of TMD
parton distributions \cite{coll2,jiyu,bjy}, whose
contribution does not vanish in the light-cone gauge,
as it happens for the standard PD functions.
Those T-odd functions can be intuitively related to
impact parameter dependent (IPD) parton distributions~\cite{burk2,burk}.
However, because of the gauge link, it is formally questionable.
For example,
in the case of the Sivers function,
it has been shown that
a model independent relation
between
this quantity and the corresponding GPD
is still to be found \cite{mgm}.

The Boer--Mulders function is being measured
through
Semi-inclusive deep inelastic scattering (SIDIS)
\cite{gunar,Bressan:2009eu,Giordano:2009hi} 
(see Ref. \cite{Barone:2008tn} for an analysis
of these data)
and through the
Drell-Yan (DY) process in hadronic colliders
\cite{Bunce:2000uv,cino,pax}.
In particular, in Ref.~\cite{Zhang:2008nu},
the Boer-Mulders function
has been recently extracted from the data of the
unpolarized $p+D$ Drell-Yan processes measured by the
E866/NuSea Collaboration at FNAL~\cite{cino}.
However,
the extraction of $h_1^{\perp}$
is very difficult, as it always involves another
chirally-odd distribution function.
As a matter of fact, in SIDIS, the Collins fragmentation function
is required~\cite{Collins}, and in DY the observable quantity is a convolution
of two $h_1^{\perp}$ belonging to the two colliding nucleons \cite{Boer:1999mm}.
Presently, the lack of accurate data
affects the quality of the extraction.

The present experimental scenario motivates
therefore
the formulation of theoretical
estimates. In principle one should perform a calculation in QCD;
however, this is presently not possible. Lacking this
possibility, it becomes relevant to perform model calculations of the
Boer-Mulders function. A few estimates exist.
In a quark-diquark model with axial vector diquarks,
it was originally found that
$h_1^\perp$ has a different sign for the $u$ and $d$ flavors
\cite{bacch}. These findings have been corrected by the Authors
of Ref. \cite{gold}, who, following a procedure established in previous papers
\cite{gamb}, demonstrated that the sign
of $h_1^\perp$
for the $u$ and $d$ flavors
turns out to be the same, assuming the spectator diquark to be
either scalar or axial vector.
This feature is in qualitative agreement
with the pattern predicted by quark helicity-flip IPD GPDs
in models \cite{alike,bp} and in lattice simulations \cite{Gockeler:2006zu}.
It has also been found in a few other model calculations:
in the MIT bag model, in its simplest
version \cite{yuan}; in a large $N_c$ analysis of TMDs
\cite{nc}; in a phenomenological parameterization
based on the quark-diquark picture \cite{rad}.

In here, the recently proposed formalism of Refs.
\cite{Courtoy:2008vi} and
\cite{Courtoy:2008dn},
used so far for the evaluation of
the Sivers function in a Constituent Quark Model (CQM)
and in the MIT bag model, respectively, will be extended
to calculate $h_1^\perp$ for the valence quarks.
In the case of the Sivers function,
within both models,
this approach
has proven to be able to reproduce
the main features of $f_{1T}^\perp$.
Similar expectations motivate the present analysis of $h_1^\perp$.
The MIT bag model calculation presented here has an additional
purpose, namely completeness. As a matter of fact,
in Ref. \cite{yuan}, the general framework for the calculation
of T-odd TMDs has been nicely set up, but
an important contribution has been nevertheless disregarded.
In Ref.~\cite{Courtoy:2008dn}, we have
reincorporated this contribution into the calculation of the Sivers function.
We found that
the results of Ref. \cite{yuan} were
incomplete and, as a consequence,
they did not fulfill the Burkardt Sum Rule. Once this missing contribution
is properly included in the calculation, the Burkardt Sum Rule
turns out to be fulfilled to a large extent \cite{Courtoy:2008dn}.
The same problem affects the
evaluation of $h_1^\perp$ in Ref. \cite{yuan},
which is therefore retaken here, along the lines
of Ref. \cite{Courtoy:2008dn}.

The paper is structured as follows. In the second section, the main
quantities of interest are introduced and the
formalism for the calculation of $h_1^\perp$ in a CQM and
in the MIT bag model is
developed (some technical details
are given in the Appendix).
The numerical results of the calculations are presented
and discussed in the following section. In
the last section, we draw conclusions from our study.
All throughout the paper, the formalism and the results are critically
compared with those previously obtained for
the Sivers function $f_{1T}^\perp$.

\section{Formalism}

The Boer-Mulders function,  $h_{1}^{\perp {\cal Q} } (x, k_T)$
\cite{Boer:1997nt},
the quantity of interest here,
is formally defined,
according to the Trento convention
\cite{trento},
for the quark of flavor ${\cal Q}$,
through the following
expression\footnote{An equivalent expression is obtained
by performing the following changes in Eq. (\ref{bm-def-op}):
$k_x \rightarrow -k_y$
and $\gamma^2 \rightarrow \gamma^1$.
Here and in the
following, $a^\pm = (a_0 \pm a_3)/ \sqrt{2}$ and $k_T=|\vec{k}_T|$.}:
\beq
h_{1}^{\perp {\cal Q}} (x, k_T) & = &
- {M \over 2 k_x}
\,\int \frac{d\xi^- d^2\vec{\xi}_T}{(2\pi)^3}
\ e^{-i(xp^+ \xi^- -\vec{k}_T\cdot \vec{\xi}_T)} \,
\nonumber\\
& \times &
{1 \over 2} \sum_{S_z=-1,1}
\langle P S_z\vert \overline \psi_{\cal Q} (\xi^-, \vec{\xi}_T)\,
{\cal L}^{\dagger}_{\vec{\xi}_T}(\infty, \xi^-)\,
\gamma^+\gamma^2\gamma_5\,{\cal L}_0(\infty,0)
\psi_{\cal Q}(0,0) \vert P S_z\rangle \, + \mbox{h.c.} \quad,\nonumber\\
\label{bm-def-op}
\eeq
where $ {\vec S} $ is the spin of the target hadron,
the normalization of the covariant spin vector is $S^2 = -1$,
$M$ is the target mass, $\psi_{\cal Q}(\xi)$ is the quark field
and the gauge link is
\beq
 {\cal L}_{\vec{\xi}_T}(\infty, \xi^-)&=& {\cal P} \mbox{exp}\left( -ig\, \int_{\xi^-}^{\infty} \, A^+(\eta^-,\vec{\xi}_T)\, d\eta^-\right)\quad,
\eeq
\\
where $g$ is the strong coupling constant.
One should notice that this definition for the gauge link
holds in covariant (non singular) gauges,
and in SIDIS processes, since the definition of T-odd TMDs
is process dependent.
As observed in Ref. \cite{brohs} for the first time,
and later in \cite{Ji:2004wu,Collins:2004nx}
using factorization theorems,
the gauge link
contains a scaling contribution which makes the T-odd TMDs non vanishing
in the Bjorken limit.

For completeness, we recall now the definition of the Sivers function,
$f_{1T}^{\perp {\cal Q} } (x, k_T)$. Taking the proton polarized
along the $y$ axis, one has
\begin{eqnarray}
f_{1T}^{\perp {\cal Q}} (x, {k_T} ) & = &
- {M \over 2k_x}
\,\int \frac{d\xi^- d^2\vec{\xi}_T}{(2\pi)^3}
\ e^{-i(xp^+ \xi^- -\vec{k}_T\cdot \vec{\xi}_T)} \,
\nonumber\\
& \times &
{1 \over 2} \sum_{S_y=-1,1} \, S_y \,
\langle P S_y\vert \overline \psi_{\cal Q} (\xi^-, \vec{\xi}_T)\,
{\cal L}^{\dagger}_{\vec{\xi}_T}(\infty, \xi^-)\,
\gamma^+\,{\cal L}_0(\infty,0)
\psi_{\cal Q}(0,0) \vert P S_y\rangle \, + \mbox{h.c.} \quad.\nonumber\\
\label{siv-def-op}
\eeq
The difference between
the two quantities is clearly seen and physically transparent in
Eqs. (\ref{bm-def-op}) and (\ref{siv-def-op}).
The function
$h_{1}^{\perp {\cal Q} } (x, k_T)$ counts the transversely
polarized quarks (as given by the Dirac Operator $\gamma^+\gamma^2\gamma_5$
in Eq. (\ref{bm-def-op})) in an unpolarized
proton (as given by the average over the proton helicity $S_z$
in Eq. (\ref{bm-def-op})),
while
$f_{1T}^{\perp {\cal Q} } (x, k_T)$
counts the unpolarized quarks (as given by the Dirac
Operator $\gamma^+$
in Eq. (\ref{siv-def-op})) in a transversly polarized
proton (as given by the explicit transverse component $S_y$
in Eq. (\ref{siv-def-op})).
If there were no gauge links, the two T-odd functions
$f_{1T}^{\perp {\cal Q} } (x, k_T)$ and
$h_{1}^{\perp {\cal Q} } (x, k_T)$
would be identically zero.
Expanding the gauge link
to the first non-trivial order, i.e.
the next to leading one, the Boer--Mulders function
$h_{1}^{\perp {\cal Q} } (x, k_T)$, Eq.
(\ref{bm-def-op}),
becomes
\beq
h_1^{\perp {\cal Q}} (x, k_T)&=&
-\frac{M}{2 k_x}\,\int \frac{d\xi^- d^2\vec{\xi}_T}{(2\pi)^3}
\ e^{-i(xp^+ \xi^- -\vec{k}_T\cdot \vec{\xi}_T)} \,\nonumber\\
&\times&
{1 \over 2} \sum_{S_z=-1,1}
\langle P S_z \vert \overline \psi_{{\cal Q} i} (\xi^-, \vec{\xi}_T)
\, \nonumber\\
&\times&
(ig)\,\int_{\xi^-}^{\infty} \, A^+_a(\eta^-,\vec{\xi}_T)\, d\eta^-
\, T^a_{ij}\, \gamma^+\gamma^2\gamma_5\,\psi_{{\cal Q} j} (0,0)
\vert P S_z \rangle
+\mbox{h.c.} \quad.
\label{bm-nlo}
\eeq
The formalism will be presented now for the evaluation of
$h_{1}^{\perp {\cal Q} } (x, k_T)$ in the MIT bag model and in the CQM.
Let us start from the former.
At the beginning, our procedure
will follow step by step the one nicely set up in Ref. \cite{yuan}
for the evaluation of T-odd TMDs in the MIT bag model.
The properly normalized fields for the quark in the bag (see also
\cite{jaffe}) are given in terms of the quark wave function in momentum
space, which read
\begin{eqnarray}
\varphi_m(\vec{k})&=&i\, \sqrt{4\pi}\, N\, R_0^3
\begin{pmatrix}
 t_0(|\vec{k}|) \chi_m
\\
{\vec{\sigma} \cdot \hat{k}}\,t_1(|\vec{k}|)\, \chi_m
\end{pmatrix} \quad,
\label{bagwf}
\end{eqnarray}
with the normalization factor $N$
\begin{eqnarray}
N&=&\left( \frac{\omega^3}{2R_0^3\, (\omega-1)\sin^2\omega}
\right)^{1/2}\quad;\nonumber
\end{eqnarray}
where $\omega=2.04$ for the lowest mode and $R_0=4\omega/M$
is the bag radius.
The functions $t_i(k)$ are defined as
\begin{eqnarray}
 t_i(k) &=&\int_0^1 u^2\, du\, j_i(ukR_0)j_i(u\omega)\quad.
\end{eqnarray}
\\
Following Ref. \cite{yuan}, we fix the other ingredients
of Eq. (\ref{bm-nlo}), having in mind Fig. 1 (and its h.c.).
In particular, the
gluon propagator is treated in a perturbative way and it is
assumed that it is not modified in the bag medium.
This leads, in the Landau gauge, to
\beq
h_{1}^{\perp {\cal Q}} (x, k_T)&=& \frac{g^2\,M\, E_P}{k_{x}}
\int \frac{d^3 k_3}{(2\pi)^3}
\frac{d^3 k_1}{(2\pi)^3}\frac{d^4 q}{(2\pi)^4}\,
\delta\left( k_1^+-xP^++q^+\right) \delta^{(2)}
\left( \vec{k}_{1T}-\vec{k}_T+\vec{q}_T\right) \nonumber\\
&\times&2\pi \delta(q_0)\, \frac{1}{q^+ + i\epsilon}\frac{1}{q^2+i\epsilon}
\nonumber\\
&\times&
{1 \over 2} \sum_{S_z=-1,1}
\sum_{\beta}\sum_{m_1,m_2,m_3, m_4} \, T^a_{ij}T^a_{kl}\,
\langle P S_z |b_{{\cal Q} m_1}^{i\dagger}b_{{\cal Q} m_2}^{j}
    b_{\beta m_3}^{k\dagger}b_{\beta m_4}^{l} |P S_z \rangle\nonumber\\
&\times&
\varphi^{\dagger}_{m_1} (\vec{k}_{1})\,\gamma^0 \gamma^+\gamma^2\gamma_5\,
 \varphi_{m_2} ( \vec{k})
 \varphi^{\dagger}_{m_3} ( \vec{k}_{3})\,\gamma^0 \gamma^+\, \varphi_{m_4}
( \vec{k}_{3}-\vec{q})
 +\mbox{h.c.}
 \quad.
\label{nlo-start}
\eeq
The last expression corresponds to the definition of the Boer-Mulders
function in the MIT bag model, given in Ref.~\cite{yuan}.

Substituting in Eq.~(\ref{nlo-start}) the identity
\beq
\frac{1}{q^+-i\epsilon}-\frac{1}{q^++i\epsilon}&=& i\,2\pi\, \delta(q^+)\quad,
\nonumber
\eeq
%
and performing the trivial integrations, one gets
\beq
h_{1}^{\perp {\cal Q}} (x, k_T)&=&-2 i g^2\,\frac{M\, E_P}{k_{x}}\,
\int \frac{d^2 \vec{q}_T}{(2\pi)^5}
\,\frac{1}{q^2}\nonumber\\
&\times&
{1 \over 2} \sum_{S_z=-1,1}
\sum_{\beta}\sum_{m_1,m_2,m_3, m_4} \, T^a_{ij}T^a_{kl}\,
\langle P S_z |b_{{\cal Q} m_1}^{i\dagger}b_{{\cal Q} m_2}^{j}
    b_{\beta m_3}^{k\dagger}b_{\beta m_4}^{l} |P S_z\rangle\nonumber\\
&\times&
\varphi^{\dagger}_{m_1} (\vec{k}-\vec{q}_T)\,\gamma^0 \gamma^+\gamma^2
\gamma_5\,
 \varphi_{m_2} (\vec{k})
\int \frac{d^3 k_3}{(2\pi)^3}
\varphi^{\dagger}_{m_3} ( \vec{k}_{3})\,\gamma^0 \gamma^+\, \varphi_{m_4}
(\vec{k}_{3}-\vec{q}_T)
 \quad.
\label{bm-bag-ready}
\eeq
The last line here depends on the four spin indices, whose combinations
are dictated by the spin-flavor-color matrix elements
given in the second line.
Let us see now where the differences between the present calculation
and that of Ref. \cite{yuan} arise.
Following the notation of Ref.~\cite{Courtoy:2008dn}, we write
\beq
\varphi_{m_1}^\dagger(\vec{k}-\vec{q}_T)\gamma^0\gamma^+\gamma^2\gamma_5\,
\varphi_{m_2}(\vec{k})&=&I_{m_1}(\vec{k}, \vec{q}_{T})
\,\delta_{m_1m_2}+J_{m_1}(\vec{k}, \vec{q}_{T})\,\delta_{m_1,-m_2}\quad,
\label{k3}
\eeq
\beq
 \int\frac{d^3k_3}{(2\pi)^3}\varphi_{m_3}^\dagger(\vec{k}_3)\gamma^0\gamma^+
\varphi_{m_4}(\vec{k}_3-\vec{q}_{T})&=&F_{m_3}(\vec{q}_{T})
\delta_{m_3m_4}+ H_{m_3}(\vec{q}_{T}) \delta_{m_3,-m_4}\quad.
\label{comb}
\eeq
In these expressions, all the possible
helicity combinations
in the initial and final states are emphasized.
To our understanding, there are no physical reasons to rule out none
of the above possible contributions.
It is easily realized that both
the terms with no-helicity flip of the quark $3$, i.e. $F_{m_3}$,
as well as the one allowing the helicity-flip of this quark, i.e. $H_{m_3}$,
are non-vanishing under the integration over $d^3 {k}_3$.
This result is in contrast with the result presented
in Ref.~\cite{yuan}, where only a term similar to $F_{m_3}$ is considered.
As we explained in Ref. \cite{Courtoy:2008dn},
this result only applies if the integral is
performed taking $\vec q_T$ along the $z$ direction. However, this
is incorrect because,
in any DIS process,
the operator structure is determined
by the direction of the virtual photon.
Here, the operator $\gamma^+$ results from having taken the
photon along the $z$ axis.
Therefore we no longer have the freedom to choose $z$
as the direction of the exchanged gluon,
which must then lie in the $(x,y)$ plane.
Besides, one can check that the integral
Eq.~(\ref{comb}) does depend on the direction of $\vec q_T$.
Moreover,
if the findings of Ref.~\cite{yuan} were correct, it
would mean that no helicity flip for the interacting
quark could occur, a restriction which does
not  have any physical motivation.
Thus the present calculation differs from the previous one in that
we take into consideration both terms of Eq.~(\ref{comb}).
By the same argument, the expression
$\varphi_{m_1}^\dagger(\vec{k}-\vec{q}_T)\gamma^0\gamma^+
\gamma^2\gamma_5
\varphi_{m_2}(\vec{k})$
in Eq. (\ref{k3})
also contains both helicity-flip and non-flip terms.

Using Eqs. (\ref{k3})
and (\ref{comb}) in Eq. (\ref{bm-bag-ready}),
the Boer-Mulders function reads
\begin{eqnarray}
h_{1}^{\perp {\cal Q}}(x,k_{T})&=& -2ig^2\frac{M E_P}{k_x}
\int \frac{d^2q_{T}}{(2\pi)^5}\frac{1}{q^2}
\nonumber\\ & \times &
{1 \over 2} \sum_{S_z=-1,1}
\sum_{\beta}\sum_{m_1,m_2,m_3,m_4} T^a_{ij}T^a_{kl}\,
\langle P S_z|  b_{{\cal Q} m_1}^{i\dagger}b_{{\cal Q} m_2}^{j}
b_{\beta m_3}^{k\dagger}b_{\beta m_4}^{l} |P S_z \rangle
\nonumber\\ & \times &
\Big\{
I_{m_1}(\vec{k}, \vec{q}_{T})\,\delta_{m_1m_2}+J_{m_1}(\vec{k}, \vec{q}_{T})\,\delta_{m_1,-m_2}
\Big\}
\Big\{
F_{m_3}(\vec{q}_{T})\,\delta_{m_3m_4}+H_{m_3}(\vec{q}_{T})\,\delta_{m_3,-m_4}
\Big\}\,.\nonumber\\
\label{bm-ijfh}
\end{eqnarray}

After the evaluation of the spin-flavor-color matrix elements,
  \beq
  \sum_{\beta} \,T^a_{ij}T^a_{kl}\,\langle P S_z|  b_{{\cal Q} m_1}^{i\dagger}b_{{\cal Q} m_2}^{j}b_{\beta m_3}^{k\dagger}b_{\beta m_4}^{l} |P  S_z\rangle&=&
C_{{\cal Q}S_z}^{m_1m_2, m_3m_4}\quad,
\label{me}
\eeq
performed assuming SU(6) symmetry,
for ${\cal Q}=u,d$ and $S_z=1(-1)$ (these coefficients
are listed in the Appendix), and
after a straightforward calculation,
one is left with the final expression
\beq
h_{1}^{\perp u(d)}(x,k_{T})&=&- g^2\,\frac{M E_P}{k^x}
\,
\int \frac{d^2q_{T}}{(2\pi)^2}\frac{C^2}{q^2}
\nonumber\\
&\times &
\left\{
6(3) \, I_1\,F_1+ 1 (2)  \, I_2\,F_2 + 1 (2)  \, J_1\,H_1 + 1 (2)
\, J_2\,H_2
\right\}\quad,
\label{bm-res-bag}
\eeq
with the eight functions $I_{1,2}, F_{1,2}, H_{1,2}, J_{1,2}$
and the $c$-number $C$ given in the Appendix.
The two first terms on the r.h.s. of~(\ref{bm-res-bag}) are the
contributions involving  quarks which do not flip helicity.
On the other hand, the last two terms on the r.h.s.
of the same expression represent the contribution
due to the double helicity flip of the quarks.
This contribution will turn out to be
non negligible. We reiterate that
the result Eq. (\ref{bm-res-bag}) is different from  that of the
previous calculation in the bag model
~\cite{yuan}, which includes only
a $F_{m_3}$-like term.

Through the coefficients Eq. (\ref{me}) it is possible to
reconstruct what happens at the level of the quark helicity
in a perfectly transparent way.
The r\^ole of each contribution can be followed and evaluated.
The dominant contributions are the non-flipping $I_1F_1$ and
the double-flipping $J_1H_1$ ones.
Due to the spin-flavor-color coefficients, i.e., due to the
SU(6) symmetry assumption, the non-flipping
term is bigger than the double-flipping contribution.
The improvement of the Boer-Mulders in the MIT bag model
with respect to the previous calculation~\cite{yuan}
quantitatively consists in the addition of the double-flipping contribution.
In effect,
the other two terms, $I_2F_2$ and $J_2H_2$, are governed by the product of the
two lower components of the bag wave function
(cf. Eqs. (A8) and (A14) in the Appendix),
which encodes the most relativistic contribution
arising in the MIT bag model.
They turn out to be a few orders of magnitude smaller
than the dominant ones, arising from the interference between
the upper and lower parts of the bag wave function.
We will see now that this happens also if a proper
non relativistic reduction (NR) of the gauge link, suitable
for CQM calculations, is performed.

The calculation scheme
for T-odd functions in a CQM
has been completely set up in Ref.~\cite{Courtoy:2008vi}, where it has
been applied
to the Sivers function.
Since
the formalism required here for the Boer-Mulders function is the same
used in our previous paper to evaluate $f_{1}^{\perp {\cal Q} } (x, k_T)$,
we refer to that paper to obtain
a workable formula for $h_{1}^{\perp {\cal Q} } (x, k_T)$.
As it has been discussed after  Eqs. (1) and (3),
with respect to the $f_{1}^{\perp {\cal Q} } (x, k_T)$ calculation,
one has to change
the polarization of the proton and the Dirac structure
of the quark operator.
Starting from the definition, Eq. (\ref{bm-def-op}),
it is found therefore that, in a CQM,
the equivalent of the expression Eq.~(\ref{bm-bag-ready}),
previously obtained in the MIT bag model, is:
\begin{eqnarray}
h_{1}^{\perp {\cal Q}} (x, {k_T} ) & = &
- 2 i g^2
{
M^2 \over k_x
}
\int
d \vec k_1
d \vec k_3
{d^2 \vec q_T \over (2 \pi)^2}
\delta(k_1^+ - xP^+)
\delta(\vec k_{1 T} + \vec q_T - \vec k_T) {\cal M}^{\cal Q}\quad,
\label{start2}
\end{eqnarray}
where
\begin{eqnarray}
{\cal M}^{\cal Q} & = &
\sum_{{\cal F}_3,m_1,c_3,m_2,c_4,m_3,i,m_4,j}
\delta_{(S_z,r,m_1,m_2,l_n,m_3,m_4,i,j,c_3,c_4)} \,\,
{1 \over 2} \sum_{S_z=-1,1}
\nonumber \\
& \times &
\Psi^{\dagger}_{r \, S_z}
\left ( \vec k_1 \{m_1,i,{\cal Q}\}; \, \vec k_3 \{m_3,c_3,{\cal F}_3 \};
\, - \vec k_3 - \vec k_1,  l_n  \right )
\nonumber \\
& \times &
T^a_{ij}  T^a_{c_3c_4}
V^h(\vec k_1, \vec k_3, \vec q)
\nonumber \\
& \times &
\Psi_{r \, S_z}
\left (\vec k_1 + \vec q, \{m_2,j,{\cal Q}\}; \, \vec k_3 -
\vec q, \{m_4,c_4,{\cal F}_3 \};
\, - \vec k_3 - \vec k_1,  l_n  \right )~,
\label{M}
\end{eqnarray}
where, $l_i=\{m_i,c_i,{\cal Q}\}$ represents the set
of helicity, color and flavor quantum numbers, respectively,
and the vector $\Psi_{r \, S_z}$ is an intrinsic proton state.
Given the diagram shown in Fig. 1,
the interaction
$
V^h(\vec k_1, \vec k_3, \vec q)
$
reads
\beq
V^h(\vec k_1, \vec k_3, \vec q)
&=& \frac{1}{q^2}\, \bar u_{m_1}\left(\vec{k}_1 \right)
\,\gamma^+\gamma^2\gamma_5\,
u_{m_2}\left(\vec{k}_1 + \vec q \right)
\,  \bar u_{m_3}\left(\vec{k}_3\right)
\,\gamma^+\,
u_{m_4}\left(\vec{k}_3-\vec{q}\right)
\quad.
\label{int}
\eeq
%
%
$\Psi_{r \, S_z}$
can be factorized into a completely
antisymmetric color wave function, $ \Gamma $,
and a symmetric spin-flavor-momentum
state, $\Phi_{sf}$, as follows:
\begin{eqnarray}
\Psi_{r \, S_z}=
\Phi_{sf,S_z}
\left ( \vec k_1 \{m_1, {\cal Q}\}; \, \vec k_3 \{m_3,{\cal F}_3 \};
\, - \vec k_1 - \vec k_3,  \{m_n, {\cal F}_n \}  \right )
\Gamma (i,c_3,c_n)~.
\label{fact}
\end{eqnarray}
The matrix element of the color operator in Eq. (\ref{M})
can be therefore immediately evaluated:
\begin{eqnarray}
\sum_{c_3,c_4,i,j}
\Gamma^\dagger(i,c_3,c_n)
T^a_{ij}  T^a_{c_3c_4}
\Gamma(j,c_4,c_n)= - { 2 \over 3 }~,
\label{chicol}
\end{eqnarray}
which is the well-known result for the exchange of
one gluon between quarks in a color singlet
3-quark state \cite{mmg}. Besides, as a consequence of
the symmetry of the state $ \Phi_{sf}$,
one can perform the calculation assuming that the
interacting quark is the one labeled ``1'', so that,
after the evaluation of the summation on the flavors ${\cal F}_3$,
${\cal M}^\alpha$ can be written, for the $u$
and $d$ flavors, as follows:
\begin{eqnarray}
{\cal M}^{u(d)} & = &
\left ( - { 2 \over 3} \right ) \cdot 3 \cdot
{1 \over 2} \sum_{S_z=-1,1}\,\,
\sum_{m_1,m_2,m_3,m_4}
\Phi_{sf,S_z}^{\dagger}
\left ( \vec k_1, m_1; \vec k_3, m_3;
\, - \vec k_1 - \vec k_3,  m_n  \right )
\nonumber
\\
& \times &
{ 1 \pm \tau_3(1) \over 2 }
V^h_{\mbox{\tiny NR}}(\vec k_1, \vec k_3, \vec q)
\nonumber
\\
& \times &
\Phi_{sf , S_z}
\left (\vec k_1 + \vec q, m_2; \, \vec k_3 -
\vec q, m_4;
\, - \vec k_1 - \vec k_3,  m_n  \right )~,
\label{Mu}
\end{eqnarray}
where the helicity $m_n$ of the spectator quark is determined
by $m_1$, $m_3$ and $S_z$ due to angular momentum conservation.
Some remarks concerning the interaction, Eq. (\ref{int}), are in order.
In general, the interaction can be separated according
to the possible helicity combinations of the interacting quarks
in the initial and final state.
Besides, since the wave functions to be used are NR, a
NR reduction of the interaction has to be
obtained. This NR reduction is given by the
$V^h_{\mbox{\tiny NR}}(\vec k_1, \vec k_3, \vec q)$ expression in Eq. (\ref{Mu}).
In order to obtain it,
we follow the procedure developed for the Sivers function
in Ref.~\cite{Courtoy:2008vi},
using therefore
the definitions of free four-spinors in Eq. (\ref{int}) and
performing a NR expansion as it is commonly done in nuclear physics.
At $O\left( {k^2 / M^2} \right)$ one obtains
\beq
V_{\mbox{\tiny NR}}^h\left( \vec{k}_1,\vec{k}_3, \vec{q}\right)
&=&V^{no-flip}_{\mbox{\tiny NR}}
\left( \vec{k}_1,\vec{k}_3, \vec{q}\right)
+
V^{double-flip}_{\mbox{\tiny NR}}
\left( \vec{k}_1,\vec{k}_3, \vec{q}\right) \quad,
\label{flipno}
\eeq
i.e., the relevant processes for the evaluation
of $h_1^\perp$ are the ones where either
the interacting and active quarks
do not flip their helicities, or they flip them both.
The expressions for the $V^{no-flip}_{\mbox{\tiny NR}}$
and
$V^{double-flip}_{\mbox{\tiny NR}}$
potentials are given in the Appendix.
As explained in Ref.~\cite{Courtoy:2008vi} in the case
of the Sivers function, we reiterate
that, in this approach, it is the interference of the upper
and lower components of the four-spinors
of the free quark states which leads to a non-vanishing
$h_1^\perp$.
Effectively, these interference terms in the interaction are
the ones that, in the MIT bag model previously described,
arise due to the wave function.
Terms of higher order would correspond to those
arising, in the MIT
bag model calculation, from the product of the lower
components of the quark wave function, a contribution
which has been found to be negligible.
The results of the relativistic calculation justifies therefore the
NR reduction of the interaction in the CQM calculation.

Eq. (\ref{start2}), with ${\cal{M}}^{u(d)}$
given by Eq. (\ref{Mu}),
provides a suitable formula
to evaluate $h_1^\perp$
using the proton state
in momentum space, $\Phi_{sf}$, described
in a CQM. We will here restrict our calculation to
a Harmonic Oscillator potential model with pure SU(6)
symmetry for the proton.
The choice of a SU(6) wave function is motivated to render
the comparison with
the previous calculation, performed in the MIT bag model
with SU(6) symmetry, more plausible.
It is worth recalling that,
as it has been stressed in Ref.~\cite{Courtoy:2008vi}, the breaking of SU(6)
in the CQM of Isgur and Karl~\cite{ik} does not change essentially
the results of the calculation. This has been obtained
also in the evaluation of standard PDs or GPDs in quark models
(see, i.e. \cite{h1}):
the essential features of a CQM calculation are obtained
in a pure SU(6) framework.
%
%
The formal expression of the proton state can be
given in terms of the sets of conjugated intrinsic coordinates
given in the Appendix. It reads
\beq
 | ^2 S_{1/2} \rangle_S&=& \frac{
 e^{-\left(k^2_{\rho} +k_{\lambda}^2\right)/\alpha^2}
 }{\pi^{3/2} \alpha^3} | \chi \rangle_S~,
\label{wf}
\eeq
where the spectroscopic notation $|^{2S+1}X_J
\rangle_t$with $t=A,M,S$ being the symmetry type, has been used,
and $    | \chi \rangle_S $ is the standard SU(6)
vector describing the spin-flavor structure of the proton.
The parameter $\alpha^2 = m \omega$ of the H.O.
potential is fixed to the value 1.35 fm$^{-2}$, in order
to reproduce the slope of the proton charge form factor at zero
momentum transfer~\cite{mmg}.

We have now all the necessary ingredients to
write the final expression for $h_1^{\perp u(d)}$,
which reads
\beq
h_1^{\perp u(d)} (x, {k_T} ) &=& g^2\frac{M^2}{k_x}
\, \left(\frac{3}{2} \right)^{3/2} \frac{1}{2\,\pi^{3/2} \alpha^3}
\int \frac{d^2\vec{q}_T}{(2\pi)^2}\, \sqrt{\frac{3}{2}}\frac{\sqrt{2}
\,k_\lambda^0}{\vert  k_\lambda^0- k_\lambda^z\vert}\,
\frac{1}{q^2} \, e^{-\frac{1}{\alpha^2}
\left[k_\lambda^2 +\frac{7}{8} q_T^2-\sqrt{\frac{3}{2}}
\vec{q}.\vec{k}_{\lambda}\right]
}
\nonumber\\
\nonumber\\
&&\hspace{4cm}
\left[
1(0) B+\frac{1}{3} \left(\frac{2}{3}  \right)
\left\{ A+\frac{q^2q^x}{48m^2}
\right\}
\right]\quad,
\label{resnr}
\eeq
with
$k_\lambda^0 = \sqrt{ m^2 +  k_\lambda^2}$, and
\beq
\vec{k}_{\lambda}=\sqrt{\frac{3}{2}}(\vec{q}-\vec{k})~,&&\quad
k_{\lambda}^z=\frac{\frac{3}{2}m^2 + \vec{k}_{\lambda T}^2
- 3 x^2 P^{+ 2} }{2 \sqrt{3}P^+ x}\nonumber\\
k_{\lambda}^2&=&k_{\lambda}^{z 2}+\vec{k}_{\lambda T}^2\quad.
\eeq
The explicit expressions for the functions $A,B$ are given in the Appendix.

\section{Results and discussion}

To evaluate numerically
Eqs.
(\ref{bm-res-bag})
and
(\ref{resnr}),
the
strong coupling constant $g$, and therefore
$\alpha_s(Q^2)$, has to be fixed.
Here, the
model-independent
prescription introduced for calculations of
PDs in quark models (see, i.e., Ref. \cite{h1}) will be used.
It consists in fixing the momentum scale of the model,
the so-called hadronic scale $\mu_0^2$,
according to the
amount of momentum carried by the valence quarks.
In the present approach, for both
the MIT bag model and the CQM,
only valence quarks contribute.
The assumption that all the gluons and sea pairs in the proton
are produced perturbatively, according to NLO evolution equations,
yields $\mu_0^2 \simeq 0.1$ GeV$^2$,
if $\Lambda_{QCD}^{NLO} \simeq 0.24$ GeV.
Therefore, one has $\alpha_s(\mu_0^2)/(4 \pi) \simeq 0.13$ \cite{h1}.
This is obtained imposing that
$\simeq 55 \% $ of the momentum
is carried by the valence quarks at a scale of 0.34 GeV$^2$,
as in
typical low-energy parameterizations (see, i.e., Ref. \cite{gluu}).

The first moments of the Sivers and Boer-Mulders functions,
i.e.  the quantities
\begin{equation}
q^{(1)}  (x)
= \int {d^2 \vec k_T}  { k_T^2 \over 2 M^2}
q(x, {k_T} )\quad,
\label{momf}
\end{equation}
where $ q = h_1^{\perp {\cal Q} }$, or
$ f_{1T}^{\perp {\cal Q} }$, are depicted in Figs.~2-5.
The results for the calculation in the MIT bag model as well as in
the CQM are represented for both functions on Fig.~2.
As already mentioned in Refs.~\cite{Courtoy:2008vi, Courtoy:2008dn},
the signs of the Sivers functions
for the $u$ and $d$ flavors, negative and positive respectively,
are in agreement with both the
theoretical and experimental knowledge.
Concerning the Boer-Mulders function, no data are available but following
the pattern predicted by quark helicity-flip IPD GPDs
\cite{alike,bp}, the same sign for the $u$ and $d$ flavors is expected.
This is found in the present approach, confirming the results of
previous estimates, such as the first MIT bag model calculation
\cite{yuan}, the quark-diquark model \cite{gold}, a large $N_c$ analysis
\cite{nc} and a phenomenological parameterization
based on the quark-diquark picture \cite{rad}.
This can be clearly seen in the lower panels of Fig.~2.
The good performances of the approach,
established in the calculation of the Sivers function,
are therefore confirmed here for the Boer-Mulders function.

For the sake of clarity, the same results are rearranged
in Fig.~3, such that one can compare, in both models,
the Sivers with the Boer-Mulders functions.
It turns out that
the shape of both functions is similar in the CQM and in the MIT bag model.
In particular, the absolute value of $h_{1}^\perp$  turns out to be
a little bigger than that of $f_{1T}^\perp$.
Also the size is not too different,
since it crucially depends on the scale of the model,
which fixes the strong coupling constant $g$ appearing
in Eqs. (\ref{bm-res-bag}) and (\ref{resnr}).
A small variation of $g$ affects dramatically the results.
We have decided to fix the hadronic scale as explained at the beginning
of this section, i.e. in a model independent way, obtaining
$\alpha_s(\mu_0^2)/(4 \pi) \simeq 0.13$.
One could have instead taken the initial scales corresponding to the ones
fixed for the two different models by the Authors proposing them, i.e.
$\alpha_s(\mu_0^2)/(4 \pi) \simeq 0.11$ for the CQM \cite{mmg} and
$\alpha_s(\mu_0^2)/(4 \pi) \simeq 0.18$ for the MIT bag \cite{thomas}, respectively.
The sizes of the results in the two different models
would have come out more similar.
Since there are no precise data available, a finer tuning of these parameters
is not relevant at this moment.

In Figs. 4 and 5, a
microscopic analysis of the contributions of different helicity-flips
of the quark interacting with the virtual photon and of
the active recoiling one is presented, for
$f_{1T}^\perp$ and
$h_{1}^\perp$, respectively.

In the case of the Sivers function, the helicity-flip can occur
either for the quark interacting with the virtual photon, or for
the active recoiling one
\cite{Courtoy:2008vi, Courtoy:2008dn}.
The latter contribution has been disregarded in Ref. \cite{yuan},
while it is seen in Fig. 4 that it is very important.
We recall that, once such a term is taken properly into account,
the Burkardt Sum Rule is fulfilled to a large extent
\cite{Courtoy:2008dn}. A similar problem occurs also for
the calculation of $h_{1}^\perp$.
In the previous section, we have shown that, in the case of this function,
contributions are found if there is no helicity-flip for the two active quarks,
or if they both flip their helicities (cf. Eq. (\ref{bm-res-bag}) and (\ref{flipno})).
The latter contribution has been once more disregarded in Ref. \cite{yuan},
while it is seen in Fig. 5 that it is sizeable.
Moreover, if the two contributions are disregraded, $f_{1T}^\perp$ for a given flavor
turns out to be proportional to
$h_{1}^\perp$ for the same flavor.
Besides,
$f_{1T}^\perp$ ($h_{1}^\perp$) for a flavor turns out to be
proportional to $f_{1T}^\perp$ ($h_{1}^\perp$) for the other flavor.

These proportionalities,
are not found once the proper helicity-flip contributions
are taken into account.
Indeed,
the reincorporated term is subtractive in
the case of the Sivers function, while it is additive
in the case of the Boer-Mulders function.
This implies that the
two T-odd functions cannot be proportional.
In any case, the proportionality
found in Ref.~\cite{yuan} has no physical motivation,
as,
in order to calculate the two T-odd functions,
one is using a two-body operator
associated with FSI and therefore one should not expect
a proportionality between the $u$ and $d$ results.
On the contrary, in the calculation of conventional
PDs, in any SU(6) model calculation, the used operators
are of one-body type and therefore the results turn out
to be proportional \cite{jaffe}.

Another interesting byproduct of the present results is that
there is basically no qualitative difference between the
results in the two models,
despite the fact that one of them is NR.
As already said in the previous section,
this has to do with the fact that
the terms which encode the most relativistic contribution
arising in the MIT bag model turn out to be a few
orders of magnitude smaller than the dominant ones.
This is seen in Eqs. (A8) and (A14)
in the Appendix at the formal level.
This gives us more confidence in
the order $O(k^2/m^2)$ of the NR expansion used.

Let us now compare our results with the other microscopic
calculation of the Boer-Mulders function,
the one in the quark-diquark model \cite{gold}.
We get the same, negative, sign for the $u$ and $d$ flavors,
which has been proven to
be related to the model independent features of IPD GPDS
\cite{alike}.
On the other hand, the relative size of the $u$ and $d$ flavors
turn out to be different (at variance with \cite{gold},
we have the result for $u$ a bit larger
than for $d$). Anyway, as far as we understand, a model
parameter in Ref. \cite{gold} can change such a relative size,
making this comparison not conclusive.
What makes our approach unique,
transparent and instructive, is that
a microscopic analysis of the helicity-flip
of the different quarks involved
is not possible in two-body models,
such as the quark-diquark model. We have seen instead that
one can learn several interesting aspects of the problem
by performing this helicity-flip analysis at the single quark level.

One relevant missing part in our analysis is the evaluation
of the pQCD evolution of the results from the scale of the model to
the experimental one. Unfortunately, to our knowledge,
the anomalous dimensions of the relevant operators have not
been calculated yet and no workable evolution equations have been provided.
Some crucial steps towards the solution of this problem have been
done \cite{cecco,stef,yuan2,kang,zhou,braun}, 
and they will be considered in future work.

Other natural developments of this calculation
are the analysis in a relativistic
three-body model, as well as the connection of T-odd TMDs
with the corresponding IPD GPDs.

\section{Conclusions}

We have applied a formalism, developed previously to calculate the Sivers function, to 
the calculation of the Boer-Mulders function with the aim of understanding the microscopic mechanisms
associated with the transverse polarization of the nucleon. The interest of the calculation lies in 
the fact that, despite the difficult experimental extraction of this function, ongoing experiments
will produce data. We have applied the formalism to two very different models of nucleon structure: a non relativistic constituent quark model 
and a relativistic MIT bag model, both of which are able to interpret well the static properties of the proton.
Our calculation is complete in the sense that it takes into account 
all possible contributions, i.e. not only the
helicity non-flip contribution, but also the helicity double flip one.

The calculated moment of the Boer-Mulders function has the same negative sign for the $u$ and $d$ flavors confirming the results of
 previous predictions and estimates. This nice outcome also happened in our previous calculation of the Sivers function using the same formalism,
 where our results satisfied the Burkardt Sum Rule and followed the general trend of previous correct estimates. The shapes of both functions,
 Sivers and Boer-Mulders, are similar in the CQM and the MIT bag model.  This is a consequence of the fact that the terms carrying the most relativistc
  contribution in the bag model are a few orders of magnitude smaller thean the main contributions. It should be stressed that, the additional terms found 
  in both of these functions in the MIT bag model calculations,  are crucial to establish this behavior and also to satisfy fundamental relations like the Burkardt Sum Rule.
  
  To conclude, we have developed a formalism that allows a straightforward evaluation of all kinds of parton distributions, which can be easily
  applied to different models of hadron structure and, in particular, to two of the most succesful ones. Moreover  the results obtained are quite model independent in structure,
  and are in agreement with fundamental principles and predictions. Our analysis has the advantage, with respect to other analyses
  in other formalisms, that it shows the microscopic separation of the different helicity-flip contributions for the different quarks involved in the process
  and, in so doing, is able to understand the microscopic structure of the transverse polarization up to evolution effects.

\vskip 1.cm

\appendix
\section{Some details of the results of the calculation}

Some details are listed below.

\subsection{MIT bag model calculation}

The matrix elements Eq. (\ref{me}) are listed below,
for ${\cal Q}=u$ and $S_z=1(-1)$,
\beq
C_{u S_z=1(-1)}^{++,++}= -\frac{5}{9} \,(0) \,,\quad
C_{u S_z=1(-1)}^{++,--}= -\frac{5}{9} \,\left(-\frac{2}{9}\right)\,,\quad
C_{u S_z=1(-1)}^{--,++}= -\frac{2}{9}\,\left(-\frac{5}{9}\right) \,,
\eeq
\beq
C_{u S_z=1(-1)}^{--,--}= 0\, \left(-\frac{5}{9} \right)
C_{u S_z=1(-1)}^{+-,+-}= 0\,\left(0\right) \,,\quad \quad
C_{u S_z=1(-1)}^{+-,-+}= \frac{1}{9}\,\left(\frac{1}{9}\right) \,,
\eeq
\beq
C_{u S_z=1(-1)}^{-+,+-}= \frac{1}{9}\,\left(\frac{1}{9}\right) \,,\quad \quad
C_{u S_z=1(-1)}^{-+,-+}= 0\,\left(0\right)\quad,
\eeq
and for ${\cal Q}=d$ and $S_z=1(-1)$,
\beq
C_{d S_z=1(-1)}^{++,++}= -\frac{1}{9} \,(0) \,,\quad
C_{d S_z=1(-1)}^{++,--}= -\frac{1}{9} \,\left(-\frac{4}{9}\right)\,,\quad
C_{d S_z=1(-1)}^{--,++}= -\frac{4}{9}\,\left(-\frac{1}{9}\right) \,,
\eeq
\beq
C_{d S_z=1(-1)}^{--,--}= 0\, \left(-\frac{1}{9} \right) \quad
C_{d S_z=1(-1)}^{+-,+-}= 0\,\left(0\right) \,,\quad \quad
C_{d S_z=1(-1)}^{+-,-+}= \frac{2}{9}\,\left(\frac{2}{9}\right) \,,
\eeq
\beq
C_{d S_z=1(-1)}^{-+,+-}= \frac{2}{9}\,\left(\frac{2}{9}\right) \,,\quad \quad
C_{d S_z=1(-1)}^{-+,-+}= 0\,\left(0\right)\quad.
\eeq

The functions entering Eq. (\ref{bm-res-bag}) are listed below:

\beq
F_1&=& \int  \,d^3\vec{k}_3\,\left\{
t_0(k_3)t_0(k'_3)+k_3^z\,\frac{t_1(k_3)}{k_3}t_0(k'_3)+k_3^z
\,\frac{t_1(k'_3)}{k'_3}t_0(k_3)
\right. \nonumber \\
& + &
\left.
\,\left( k_3^2-\vec{k}_3\cdot\vec{q}_{T}\right)\frac{t_1(k_3)}{k_3}\frac{t_1(k'_3)}{k'_3}
\right\}\quad,\\
%
%
F_2&=& \int  \,d^3\vec{k}_3\,\left\{
\left(\vec{q}_T\times \vec{k}_3\right)^z\frac{t_1(k_3)}{k_3}\frac{t_1(k'_3)}{k'_3}
\right\}\quad,\\
%
%
H_1&=& \int  \,d^3\vec{k}_3\,\left\{
k_3^x\,\frac{t_1(k_3)}{k_3}t_0(k'_3)-(k_3^{x}-q^x)\,\frac{t_1(k'_3)}{k'_3}t_0(k_3)-\,\left( \vec{q}_T\times \vec{k}_3\right)^y\frac{t_1(k_3)}{k_3}\frac{t_1(k'_3)}{k'_3}
\right\}\quad,
\nonumber\\
\\
%
%
H_2&=& - \int  \,d^3\vec{k}_3\,\left\{
k_3^y\,\frac{t_1(k_3)}{k_3}t_0(k'_3)-(k_3^{y}-q^y)\,\frac{t_1(k'_3)}{k'_3}t_0(k_3)+\,\left( \vec{q}_T\times \vec{k}_3\right)^x\frac{t_1(k_3)}{k_3}\frac{t_1(k'_3)}{k'_3}
\right\}\quad,\nonumber\\
\eeq
and
\beq
I_1&=& k^x\frac{t_1(k)}{k}t_0(k')-(k^{x}-q^x)\,\frac{t_1(k')}{k'}t_0(k)-\,\left( k^{'x}k^z-k^{'z}k^x\right)\frac{t_1(k)}{k}\frac{t_1(k')}{k'}\quad,
\\
I_2&=& k^y\frac{t_1(k)}{k}t_0(k')+(k^{y}-q^y)\,\frac{t_1(k')}{k'}t_0(k)+\,\left( k^{'y}k^z+k^{'z}k^y\right)\,\frac{t_1(k)}{k}\frac{t_1(k')}{k'}\quad,\nonumber\\
\\
J_1&=& t_0(k)t_0(k')+k^z\,\frac{t_1(k)}{k}t_0(k')+k^z\,
\frac{t_1(k')}{k'}t_0(k)
\nonumber \\
& + & \,\left( k^{x 2}-q^xk^x-k^{y 2}+q^yk^y
+k^{z 2}\right)\frac{t_1(k)}{k}\frac{t_1(k')}{k'}\quad,
\\
J_2&=&-\left(2k^xk^y-q^xk^y-q^yk^x \right)\frac{t_1(k)}{k}
\frac{t_1(k')}{k'}\quad,
\eeq
with $k=|\vec{k}|$ and $\vec{k}'=\vec{k}-\vec{q}_T$.

The constant $C$ entering Eq. (\ref{bm-res-bag}) is:

\beq
C=\sqrt{\frac{2}{9}} \frac{C'}{(2\pi)^3}~,
\eeq
with
\beq
C' = 4\pi \, N^2\, R_0^6=\frac{16\omega^4}{\pi^2\, j_0^2(\omega)(\omega-1)}\frac{(2\pi)^3}{M_P^3}~,
\eeq
determined by the MIT bag model parameters.

\subsection{CQM calculation}

The potentials corresponding to gluon exchanges without quark
helicity flips or with double quark helicity flips,
Eq. (\ref{flipno}), are listed below:

\beq
V^{no-flip}\left( \vec{k}_3,\vec{k}_1, \vec{q}\right)
&=&\frac{1}{2q^2}
\left \{
-i \left(1+\frac{k_1^z}{m}+\frac{\vec{q}\cdot\vec{k}_1}{4m^2}\right)
\left(\frac{q^x}{2m}+\frac{q^xk_3^z}{4m^2}\right)\,
%
\right.
\nonumber\\
&&\left.
+\left(1+\frac{k_1^z}{m}+\frac{\vec{q}\cdot\vec{k}_1}{4m^2}\right)
\left(-\frac{k_3^y}{m}+\frac{q^y}{2m}+\frac{q^yk_3^z}{4m^2}\right)
\,\sigma_3^z
\right.
\nonumber\\
&&\left.
+\frac{1}{4m^2} \left(k_1^xq^y-q^xk_1^y \right)
\left(\frac{q^x}{2m}+\frac{q^xk_3^z}{4m^2}\right)\,
\sigma_1^z
\right.
\nonumber\\
&&\left.
+\frac{i}{4m^2} \left(k_1^xq^y-q^xk_1^y \right)  \left(-\frac{k_3^y}{m}+\frac{q^y}{2m}+\frac{q^yk_3^z}{4m^2}\right)\sigma_3^z\sigma_1^z
\right\}\quad.
\eeq
\beq
V^{double-flip}\left( \vec{k}_3,\vec{k}_1, \vec{q}\right)
&=&\frac{1}{2q^2}
\left \{
-\frac{i}{4m^2}\left( \frac{q^y}{2m}+\frac{q^yk_1^z}{4m^2}\right)\left( k_3^xq^y+k_3^yq^x\right) \sigma_3^x\sigma_1^x
\right.
\nonumber\\
&&\left.
-i \left( \frac{q^y}{2m}+\frac{q^yk_1^z}{4m^2}\right) \left( 1+\frac{k_3^z}{m}+\frac{q^yk_3^y-q^xk_3^x}{4m^2}\right)
\sigma_3^y\sigma_1^x
\right.
\nonumber\\
&&\left.
+\frac{i}{4m^2}\,  \left( \frac{q^x}{2m}+\frac{q^xk_1^z}{4m^2}\right)\left( k_3^xq^y+k_3^yq^x\right) \sigma_3^x\sigma_1^y
\right.
\nonumber\\
&&\left.
+i  \,\left( \frac{q^x}{2m}+\frac{q^xk_1^z}{4m^2}\right)\left( 1+\frac{k_3^z}{m}+\frac{q^yk_3^y-q^xk_3^x}{4m^2}\right)
\sigma_3^y\sigma_1^y
\right \}
\eeq

The set of conjugated intrinsic coordinates
used in Eq.~(\ref{wf}) is
\begin{eqnarray}
\vec R = { 1 \over \sqrt{3} } ( \vec{r_1} + \vec{r_2} + \vec{r_3} )
& \leftrightarrow &
\vec K = { 1 \over \sqrt 3} ( \vec{k_1} + \vec{k_2} + \vec{k_3} )
\nonumber ~,
\\
\vec \rho = { 1 \over \sqrt 2} ( \vec{r_3}  - \vec{r_2} )
& \leftrightarrow &
\vec{k_{\rho}} = { 1 \over \sqrt 2} ( \vec{k_3} -  \vec{k_2} )
\nonumber ~,
\\
\vec \lambda = { 1 \over \sqrt 6} ( \vec{r_2} + \vec{r_3} - 2 \vec{r_1} )
& \leftrightarrow &
\vec{k_{\lambda}} = { 1 \over \sqrt 6} ( \vec{k_2} + \vec{k_3} - 2 \vec{k_1} )
~.
\label{coor}
\end{eqnarray}
\\
The $B$ and $A$ functions appearing in
Eq. (\ref{resnr}) are
\beq
B&=& -\frac{q^x}{2m}\left(
1+\frac{k_{\rho}^z}{2\sqrt{2}}-\frac{3}{\sqrt{6}}\frac{\bar k_{\lambda}^z}{2m}-\frac{q_x^2+q_y^2-q_xk_x-q_yk_y}{4m^2}
\right)
\nonumber\\
A&=&-\frac{5}{6}\frac{q_x}{2m}+\frac{3}{\sqrt{6}}\frac{\bar k_{\lambda}^zq_x}{4m^2}+\frac{1}{24m^3}\left(
4q_x^3-4q_x^2 k_x+4q_xq_y^2-q_xq_y^2-q_xq_yk_y-3k_xq_y^2
\right)
\eeq

\acknowledgments
\vskip 2mm
We are grateful to S. Noguera and
G. Goldstein for
fruitful discussions, and
J.C. Peng and A. Prokudin who suggested us to start this investigation.
S.S. thanks the Department of Theoretical Physics of the Valencia
University for warm hospitality;
A.C. and V.V. thank the Department of Physics of the Perugia
University for warm hospitality.  
This work is supported in part by the INFN-MEC agreement FPA2008-03646-E/INFN; by HadronPhysics2, 
a FP7-Integrating Activities and Infrastructure Program of the European Commission, under
Grant 227431;
by the MICINN (Spain) grant FPA2007-65748-C02-01 and the grant AP2005-5331.

\newpage

\section*{Figure Captions}

\vspace{1em}\noindent
{\bf Fig. 1}:
The contributions to the Sivers and Boer-Mulders
functions in the present approach. The graph has been
drawn using JaxoDraw \cite{Binosi:2003yf}.

\vspace{1em}\noindent
{\bf Fig. 2}:
Upper (lower) panel, left: the first moment of the
Sivers (Boer-Mulders) function, Eq. (\ref{momf}),
for the $u$ flavor, in the NR (full) and MIT bag
models (dashed); right: the same for the $d$ flavor.

\vspace{1em}\noindent
{\bf Fig. 3}:
Left panel: the first moment for the
Sivers and Boer-Mulders functions in the NR model.
Dashed curve: the Sivers function for $d$;
dot-dashed curve: the Sivers function for $u$;
long dashed curve: the Boer-Mulders function for $d$;
full curve: the Boer-Mulders function for $u$.
Right panel: the same in the MIT Bag model.

\vspace{1em}\noindent
{\bf Fig. 4}:
Comparison between the results for the first moment of the
Sivers function in the NR model
and in the MIT bag model.
Left panel: results in the NR model.
Full (Long dashed) curve: full result for the flavor $u$ ($d$);
dot dashed (dashed) curve:  result with the spin-flip occurring only
for the interacting $u$ ($d$) quark (as in Ref. \cite{yuan}).
Right panel: the same for the MIT Bag model

\vspace{1em}\noindent
{\bf Fig. 5}:
The same of Fig. 4, but for the Boer-Mulders function.

\newpage

\begin{figure}[ht]
\includegraphics{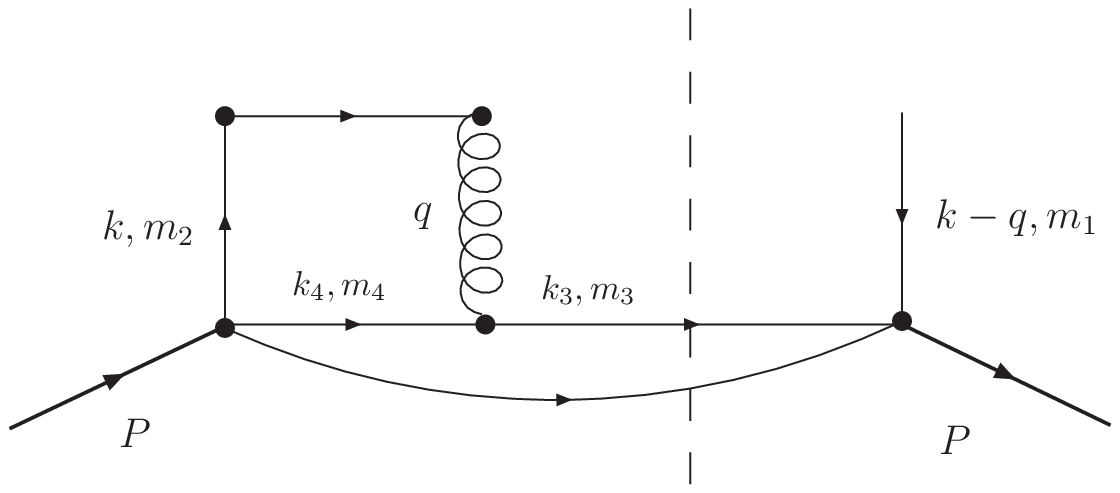}
\vspace{12.0cm}
\caption{}
\end{figure}

\newpage

\begin{figure}[ht]
\includegraphics{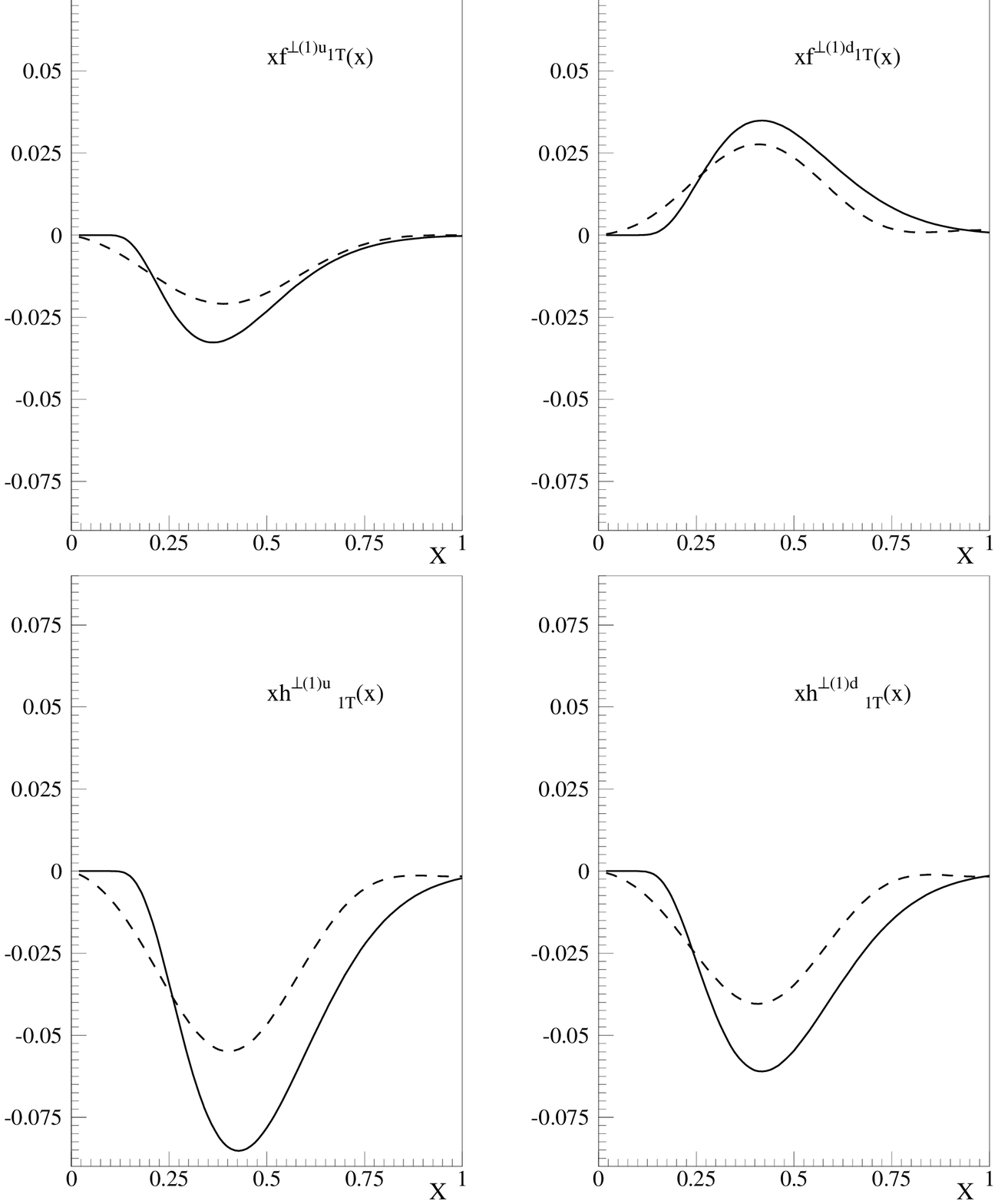}
\vspace{22.0cm}
\caption{}
\label{filippo}
\end{figure}

\newpage

\begin{figure}[h]
\includegraphics{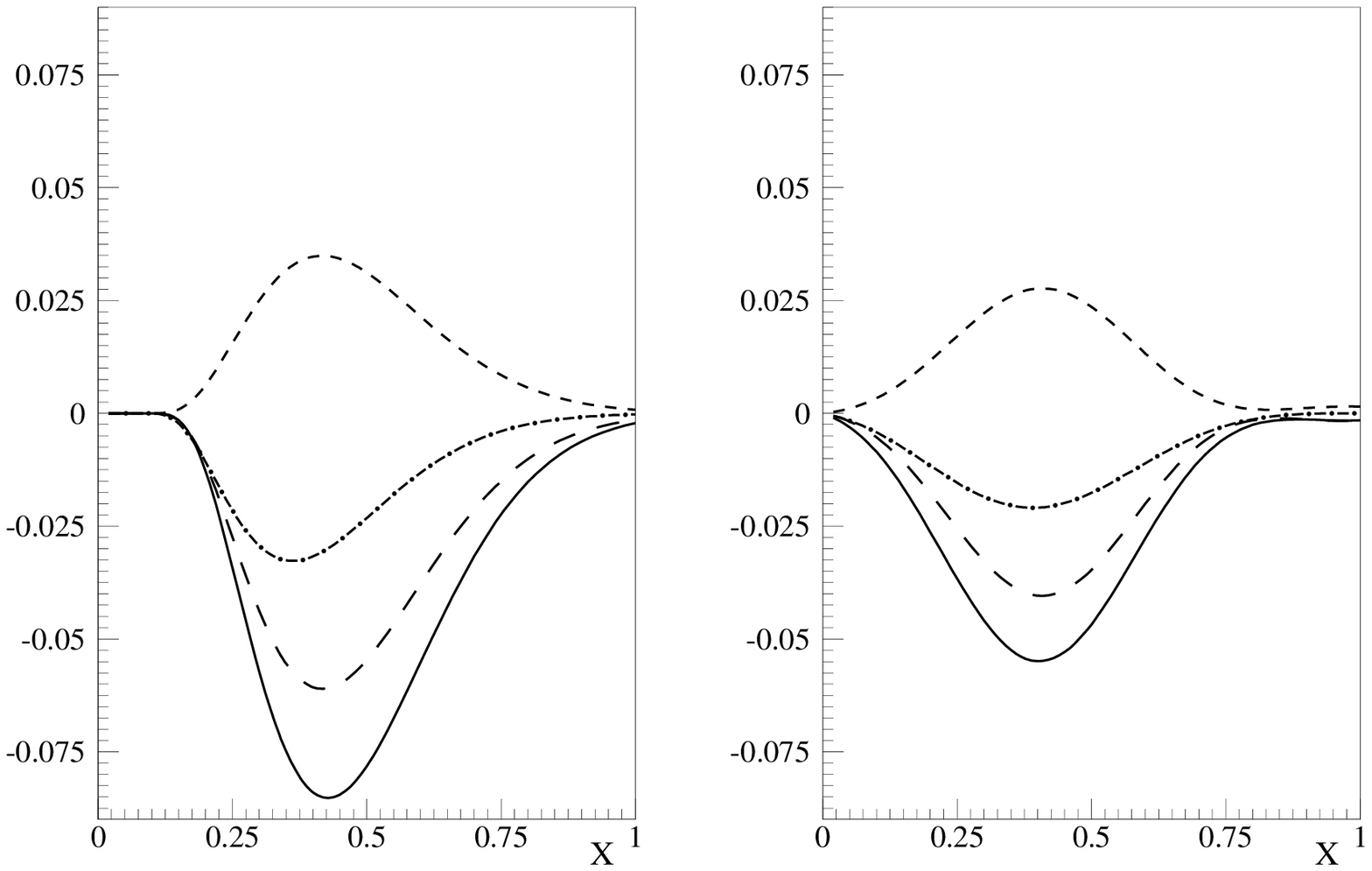}
\vspace{15.0cm}
\caption{}
\end{figure}

\newpage

\begin{figure}[h]
\includegraphics{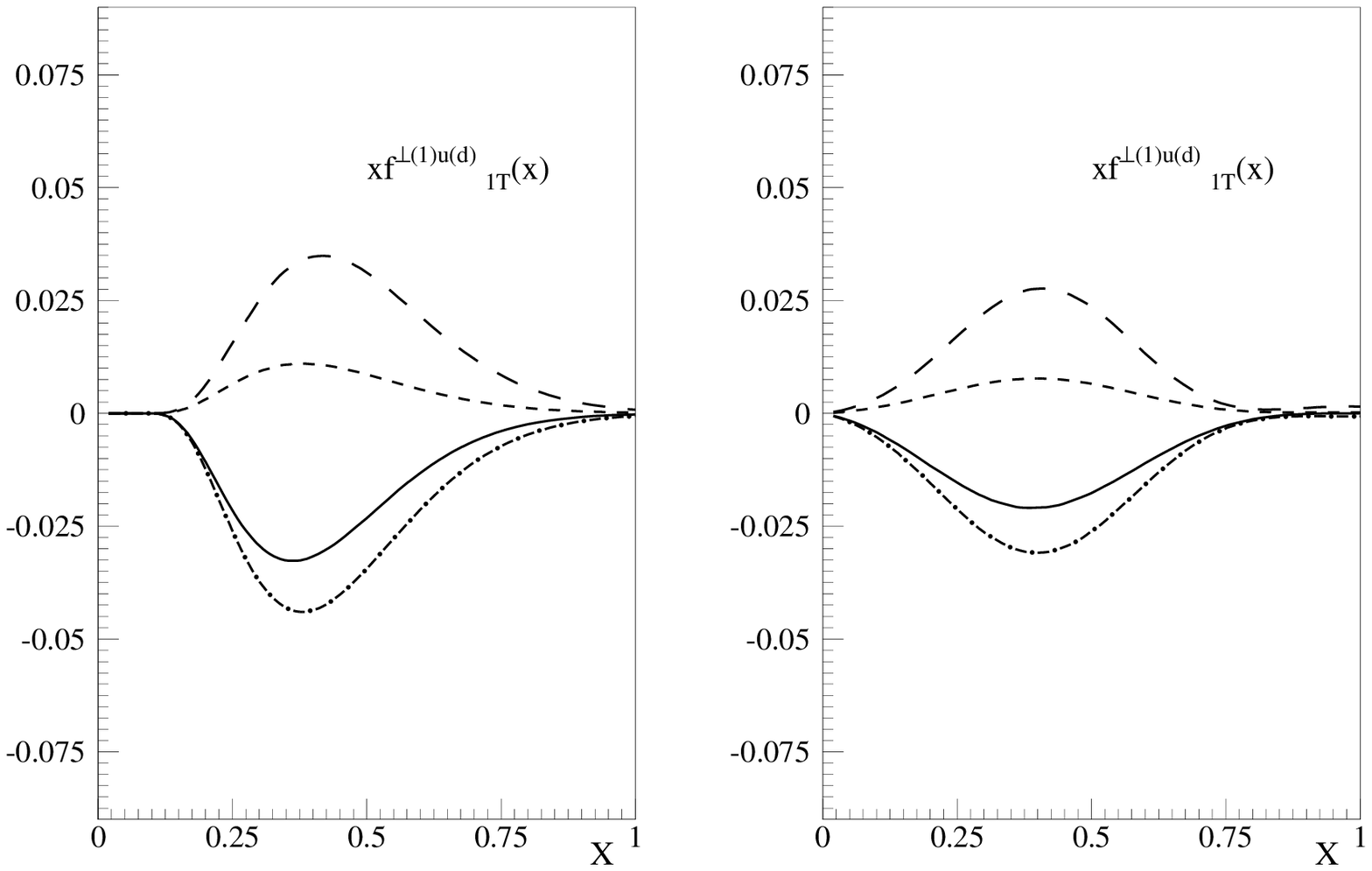}
\vspace{15.0cm}
\caption{}
\end{figure}

\newpage

\begin{figure}[h]
\includegraphics{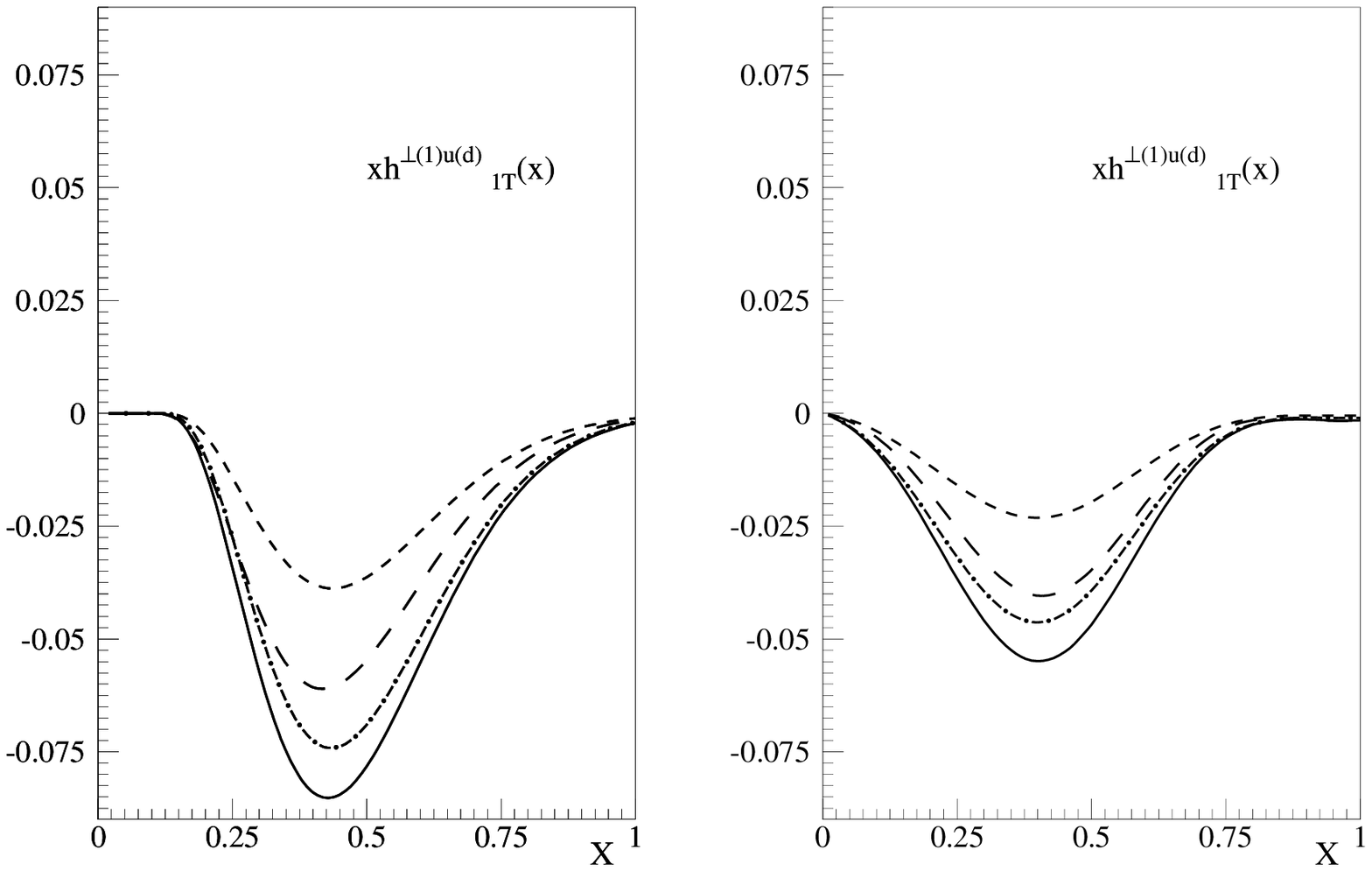}
\vspace{15.0cm}
\caption{}
\end{figure}


\begin{thebibliography}{13}


\bibitem{bdr}
  V.~Barone, A.~Drago and P.~G.~Ratcliffe,
  Phys.\ Rept.\  {\bf 359} (2002) 1
  [arXiv:hep-ph/0104283].

\bibitem{ferrara}
Proceedings of ``Transversity 2008'', May 2008, Ferrara, Italy, 2009
{\it in press}.


\bibitem{Boer:1997nt}
  D.~Boer and P.~J.~Mulders,
  Phys.\ Rev.\  D {\bf 57} (1998) 5780
  [arXiv:hep-ph/9711485].

\bibitem{Courtoy:2008vi}
  A.~Courtoy, F.~Fratini, S.~Scopetta and V.~Vento,
  Phys.\ Rev.\  D {\bf 78} (2008) 034002
  [arXiv:0801.4347 [hep-ph]].

\bibitem{Courtoy:2008dn}
  A.~Courtoy, S.~Scopetta and V.~Vento,
  Phys.\ Rev.\  D {\bf 79}, 074001 (2009)
  [arXiv:0811.1191 [hep-ph]].


\bibitem{sivers}
  D.~W.~Sivers,
  Phys.\ Rev.\  D {\bf 41}, 83 (1990),
  Phys.\ Rev.\  D {\bf 43}, 261 (1991).

\bibitem{Burkardt:2004ur}
  M.~Burkardt,
  Phys.\ Rev.\  D {\bf 69} (2004) 091501;
  Phys.\ Rev.\  D {\bf 69} (2004) 057501~.


\bibitem{ans}
M.~Anselmino, M.~Boglione, U.~D'Alesio, A.~Kotzinian,
F.~Murgia and A.~Prokudin,
  Phys.\ Rev.\  D {\bf 71}, 074006 (2005)
  [arXiv:hep-ph/0501196],
  M.~Anselmino, M.~Boglione, U.~D'Alesio, A.~Kotzinian,
F.~Murgia and A.~Prokudin,
  Phys.\ Rev.\  D {\bf 72}, 094007 (2005)
  [Erratum-ibid.\  D {\bf 72}, 099903 (2005)]
  [arXiv:hep-ph/0507181].
  A.~V.~Efremov, K.~Goeke, S.~Menzel, A.~Metz and P.~Schweitzer,
  Phys.\ Lett.\  B {\bf 612}, 233 (2005)
  [arXiv:hep-ph/0412353],
 J.~C.~Collins, A.~V.~Efremov, K.~Goeke, S.~Menzel, A.~Metz and P.~Schweitzer,
  Phys.\ Rev.\  D {\bf 73}, 014021 (2006)
  [arXiv:hep-ph/0509076]
  W.~Vogelsang and F.~Yuan,
  Phys.\ Rev.\  D {\bf 72}, 054028 (2005)
  [arXiv:hep-ph/0507266].


\bibitem{brohs}
  S.~J.~Brodsky, D.~S.~Hwang and I.~Schmidt,
  Phys.\ Lett.\  B {\bf 530}, 99 (2002)
  [arXiv:hep-ph/0201296].

\bibitem{brodhoy}
  S.~J.~Brodsky, P.~Hoyer, N.~Marchal, S.~Peigne and F.~Sannino,
  Phys.\ Rev.\  D {\bf 65}, 114025 (2002)
  [arXiv:hep-ph/0104291].


\bibitem{coll2}
  J.~C.~Collins,
  Phys.\ Lett.\  B {\bf 536}, 43 (2002)
  [arXiv:hep-ph/0204004].


\bibitem{jiyu}
  X.~d.~Ji and F.~Yuan,
  Phys.\ Lett.\  B {\bf 543}, 66 (2002)
  [arXiv:hep-ph/0206057].


\bibitem{bjy}
  A.~V.~Belitsky, X.~Ji and F.~Yuan,
  Nucl.\ Phys.\  B {\bf 656}, 165 (2003)
  [arXiv:hep-ph/0208038].


\bibitem{burk}
  M.~Burkardt,
  Nucl.\ Phys.\  A {\bf 735}, 185 (2004)
  [arXiv:hep-ph/0302144]; Phys.\ Rev.\  D {\bf 66}, 114005 (2002)
  [arXiv:hep-ph/0209179];M.~Burkardt and D.~S.~Hwang,
  Phys.\ Rev.\  D {\bf 69}, 074032 (2004)
  [arXiv:hep-ph/0309072].


\bibitem{burk2}
  M.~Burkardt,
  Int.\ J.\ Mod.\ Phys.\  A {\bf 18}, 173 (2003)
  [arXiv:hep-ph/0207047].


\bibitem{mgm}
  S.~Meissner, A.~Metz and K.~Goeke,
  Phys.\ Rev.\  D {\bf 76}, 034002 (2007)
  [arXiv:hep-ph/0703176].


\bibitem{gunar}
 G.~Schnell,
 Talk at the CLAS12 European Workshop  	
February 25-28, 2009- Genova, Italy


\bibitem{Bressan:2009eu}
  A.~Bressan and f.~t.~C.~Collaboration,
  arXiv:0907.5511 [hep-ex].

\bibitem{Giordano:2009hi}
  F.~Giordano and R.~Lamb  [On behalf of the HERMES Collaboration],
section at
  AIP Conf.\ Proc.\  {\bf 1149} (2009) 423
  [arXiv:0901.2438 [hep-ex]].

\bibitem{Barone:2008tn}
  V.~Barone, A.~Prokudin and B.~Q.~Ma,
  Phys.\ Rev.\  D {\bf 78} (2008) 045022
  [arXiv:0804.3024 [hep-ph]].

\bibitem{Bunce:2000uv}
  G.~Bunce, N.~Saito, J.~Soffer and W.~Vogelsang,
  Ann.\ Rev.\ Nucl.\ Part.\ Sci.\  {\bf 50} (2000) 525
  [arXiv:hep-ph/0007218].

\bibitem{cino}
L.Y. Zhu {\it et al.}, [FNAL-E866/NuSea], Phys. Rev. Lett. {\bf 99},
082301 (2007); Phys. Rev. Lett. 102, 182001, (2009); 
J.C. Peng,
Proceedings of the $6^{th}$ International Conference on Perspectives
in Hadronic Physics, Trieste, Italy, May 12-16 2008,
AIP Conference Proceedings, vol. 1056, 452 (2008).

\bibitem{pax}
  V.~Barone {\it et al.}  [PAX Collaboration],
  arXiv:hep-ex/0505054.


\bibitem{Zhang:2008nu}
  B.~Zhang, Z.~Lu, B.~Q.~Ma and I.~Schmidt,
  Phys.\ Rev.\  D {\bf 77} (2008) 054011
  [arXiv:0803.1692 [hep-ph]].




\bibitem{Collins}
  J.~C.~Collins,
  Nucl.\ Phys.\  B {\bf 396}, 161 (1993)
  [arXiv:hep-ph/9208213].

\bibitem{Boer:1999mm}
  D.~Boer,
  Phys.\ Rev.\  D {\bf 60} (1999) 014012
  [arXiv:hep-ph/9902255].


\bibitem{bacch}
  A.~Bacchetta, A.~Schaefer and J.~J.~Yang,
  Phys.\ Lett.\  B {\bf 578}, 109 (2004)
  [arXiv:hep-ph/0309246].


\bibitem{gold} L.P. Gamberg. G.R. Goldstein, and M. Schlegel,
Phys. Rev. D {\bf 77}, 094016 (2008).

\bibitem{gamb} L.P. Gamberg. G.R. Goldstein, and K.A. Oganessyan,
Phys. Rev. D {\bf 67}, 071504 (2003);
Phys. Rev. D {\bf 68}, 051501 (2003).

\bibitem{alike} M. Burkardt and B. Hannafious, Phys. Lett. B {\bf 658},
130 (2008).

\bibitem{bp} B. Pasquini and S. Boffi,
Phys. Lett. B {\bf 653}, 23 (2007);
e-Print: arXiv:0705.4345 [hep-ph]

\bibitem{Gockeler:2006zu}
  M.~Gockeler {\it et al.}  [QCDSF Collaboration and UKQCD Collaboration],
  Phys.\ Rev.\ Lett.\  {\bf 98} (2007) 222001
  [arXiv:hep-lat/0612032].

\bibitem{yuan}
  F.~Yuan,
  Phys.\ Lett.\  B {\bf 575}, 45 (2003)
  [arXiv:hep-ph/0308157].

\bibitem{nc}
  P.~V.~Pobylitsa,
  arXiv:hep-ph/0301236.

\bibitem{rad} A. Bacchetta, F. Conti, and M. Radici,
Phys. Rev. D {\bf 78}, 074010 (2008).


\bibitem{trento}
  A.~Bacchetta, U.~D'Alesio, M.~Diehl and C.~A.~Miller,
  Phys.\ Rev.\  D {\bf 70}, 117504 (2004)
  [arXiv:hep-ph/0410050].


\bibitem{Ji:2004wu}
  X.~d.~Ji, J.~p.~Ma and F.~Yuan,
  Phys.\ Rev.\  D {\bf 71} (2005) 034005
  [arXiv:hep-ph/0404183].

\bibitem{Collins:2004nx}
  J.~C.~Collins and A.~Metz,
  Phys.\ Rev.\ Lett.\  {\bf 93}, 252001 (2004)
  [arXiv:hep-ph/0408249].

\bibitem{jaffe}
  R.~L.~Jaffe,
  Phys.\ Rev.\  D {\bf 11} (1975) 1953.

\bibitem{mmg}
  M.~M.~Giannini,
  Rept.\ Prog.\ Phys.\  {\bf 54}, 453 (1990).

\bibitem{ik}
  N.~Isgur and G.~Karl,
  Phys.\ Rev.\  D {\bf 18}, 4187 (1978);
  Phys.\ Rev.\  D {\bf 19}, 2653 (1979)
  [Erratum-ibid.\  D {\bf 23}, 817 (1981)].

\bibitem{h1}
  S.~Scopetta and V.~Vento,
  Phys.\ Lett.\  B {\bf 424}, 25 (1998)
  [arXiv:hep-ph/9706413].
  S.~Scopetta and V.~Vento,
  Phys.\ Lett.\  B {\bf 460}, 8 (1999)
  [Erratum-ibid.\  B {\bf 474}, 235 (2000)]
  [arXiv:hep-ph/9901324].
  M.~Traini, A.~Mair, A.~Zambarda and V.~Vento,
  Nucl.\ Phys.\  A {\bf 614}, 472 (1997).
  S.~Scopetta and V.~Vento,
  Eur.\ Phys.\ J.\  A {\bf 16}, 527 (2003)
  [arXiv:hep-ph/0201265].

\bibitem{thomas}
A.W. Thomas and W. Weise, ``The structure of the nucleon'',
Wiley-VCH, Berlin, 2001.

\bibitem{gluu}
  M.~Gluck, E.~Reya and A.~Vogt,
  Eur.\ Phys.\ J.\  C {\bf 5}, 461 (1998)
  [arXiv:hep-ph/9806404].


\bibitem{cecco}
  F.~A.~Ceccopieri and L.~Trentadue,
  Phys.\ Lett.\  B {\bf 636}, 310 (2006)
  [arXiv:hep-ph/0512372];
  arXiv:0706.4242 [hep-ph].

\bibitem{stef}
  I.~O.~Cherednikov and N.~G.~Stefanis,
  Nucl.\ Phys.\  B {\bf 802} (2008) 146
  [arXiv:0802.2821 [hep-ph]].

\bibitem{yuan2}
  W.~Vogelsang and F.~Yuan,
  Phys.\ Rev.\  D {\bf 79}, 094010 (2009)
  [arXiv:0904.0410 [hep-ph]].

\bibitem{kang} Z-B. Kang, J-W. Qiu, Phys. Rev. D79, 016003, (2009). 

\bibitem{zhou} J. Zhou, F. Yuan, Z-T Liang, Phys. Rev. D79, 114022, (2009). 

\bibitem{braun} 
V.M. Braun, A.N. Manashov, B. Pirnay, e-Print: arXiv:0909.3410 [hep-ph]. 

\bibitem{Binosi:2003yf}
  D.~Binosi and L.~Theussl,
  Comput.\ Phys.\ Commun.\  {\bf 161} (2004) 76
  [arXiv:hep-ph/0309015].









\end{thebibliography}
\end{document}